\pdfoutput=1

\documentclass[10pt,journal,compsoc]{IEEEtran}
\ifCLASSOPTIONcompsoc
  \usepackage[nocompress]{cite}
\else
  \usepackage{cite}
\fi
\ifCLASSINFOpdf
   \usepackage[pdftex]{graphicx}
  \DeclareGraphicsExtensions{.pdf,.png,.jpg,.jpeg}
\else

  \usepackage{graphicx}                
  \DeclareGraphicsExtensions{.eps} 
\fi

\graphicspath{{figures/}{images/}{./}} %

\usepackage{microtype}                 
\PassOptionsToPackage{warn}{textcomp}  
\usepackage{textcomp}                  
\usepackage{mathptmx}                  
\usepackage{times}                     
\usepackage{cite}                      
\usepackage{tabu}                      
\usepackage{booktabs}                  

\usepackage{amsmath}
\usepackage{multirow} 
\usepackage{colortbl}
\usepackage{bigdelim}
\usepackage{fancyhdr}

\usepackage{hyperref}  
\hypersetup{pdfborder={0 0 0} }  

\usepackage{enumitem}
\usepackage{amsmath}
\usepackage{setspace}

\usepackage{color}

\definecolor{myred}{RGB}{0,0,0}

\begin{document}
%
\title{Evaluating Effects of Background Stories on Graph Perception}
%
%
%
%

\author{Ying Zhao, Jingcheng Shi, Jiawei Liu, Jian Zhao, Fangfang Zhou, Wenzhi Zhang\\ Kangyi Chen, Xin Zhao, Chunyao Zhu and Wei Chen
\IEEEcompsocitemizethanks{\IEEEcompsocthanksitem Ying Zhao, Jingcheng Shi, Jiawei Liu, Fangfang Zhou, Wenzhi Zhang, Kangyi Chen, Xin Zhao and Chunyao Zhu are with the School of Computer Science and Engineering, Central South University. E-mail: zhaoying@csu.edu.cn, \{429433693, 870656034\}@qq.com, zff@csu.edu.cn, \{1093894600, 2500150552, 1064253658, 240427611\}@qq.com.
\IEEEcompsocthanksitem Jian Zhao is with the School of Computer Science, University of Waterloo. E-mail:
jianzhao@uwaterloo.ca.
\IEEEcompsocthanksitem Wei Chen is with the State Key Lab of CAD \& CG, Zhejiang University. E-mail:
chenwei@cad.zju.edu.cn.
\IEEEcompsocthanksitem Fangfang Zhou is the corresponding author.}
}

%
%

\markboth{This manuscript has been accepted by IEEE TVCG (08/2021)}%
{Shell \MakeLowercase{\textit{et al.}}: Evaluating Effects of Background Stories on Graph Perception}
%



\IEEEtitleabstractindextext{%
\begin{abstract}
 A graph is an abstract model that represents relations among entities, for example, the interactions between characters in a novel. A background story endows entities and relations with real-world meanings and describes the semantics and context of the abstract model, for example, the actual story that the novel presents. Considering practical experience and prior research, human viewers who are familiar with the background story of a graph and those who do not know the background story may perceive the same graph differently. However, no previous research has adequately addressed this problem. This research paper thus presents an evaluation that investigated the effects of background stories on graph perception. Three hypotheses that focused on the role of visual focus areas, graph structure identification, and mental model formation on graph perception were formulated and guided three controlled experiments that evaluated the hypotheses using real-world graphs with background stories. An analysis of the resulting experimental data, which compared the performance of participants who read and did not read the background stories, obtained a set of instructive findings. First, having knowledge about a graph's background story influences participants' focus areas during interactive graph explorations. Second, such knowledge significantly affects one's ability to identify community structures but not high degree and bridge structures. Third, this knowledge influences graph recognition under blurred visual conditions. These findings can bring new considerations to the design of storytelling visualizations and interactive graph explorations.
\end{abstract}

}

\maketitle

\IEEEdisplaynontitleabstractindextext

%
\IEEEpeerreviewmaketitle

\IEEEraisesectionheading{\section{Introduction}\label{sec:introduction}}

\begin{figure}[!ht]
	\centering
	\vspace{0.05cm}  
	\setlength{\abovecaptionskip}{0.1cm}   
	\setlength{\belowcaptionskip}{-0.1cm}   
	\includegraphics[width=\columnwidth]{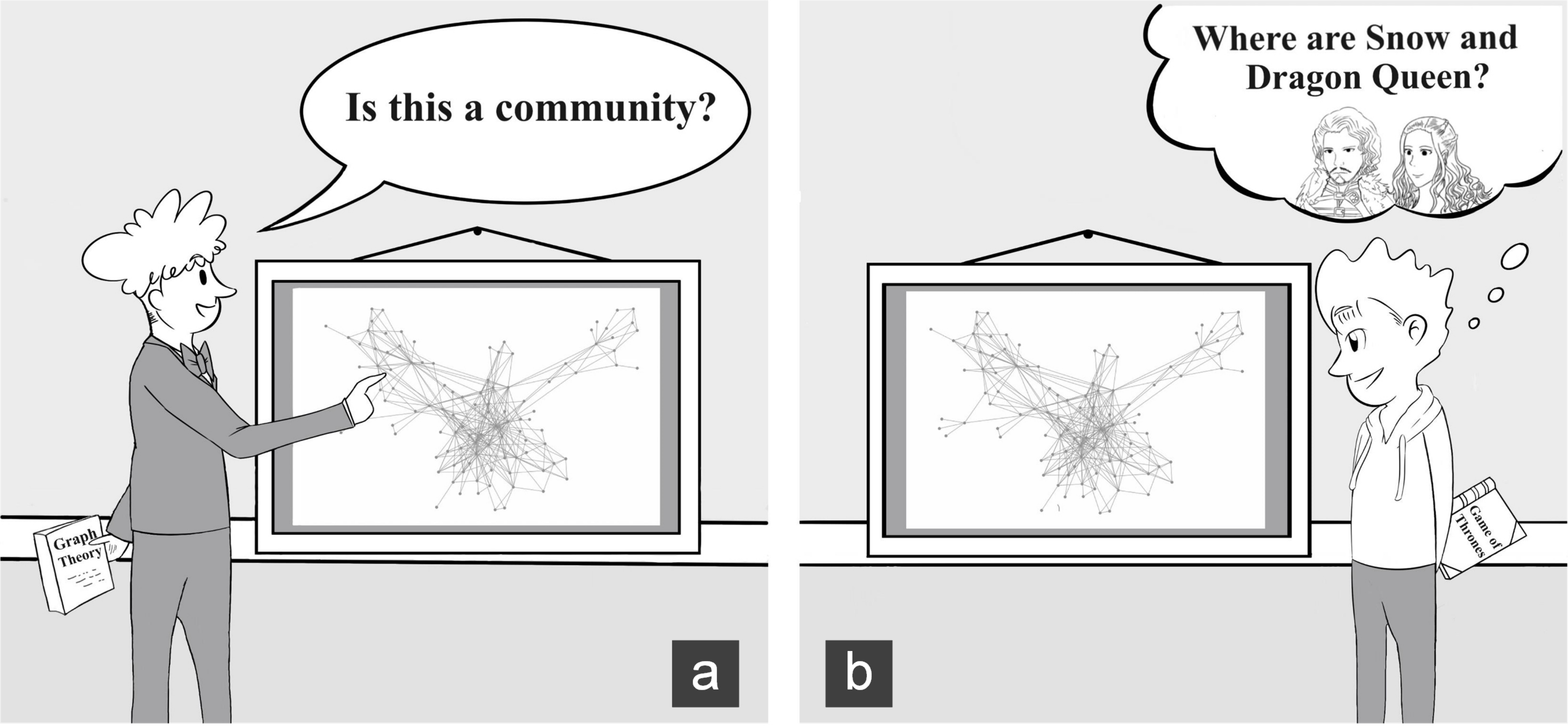}
	\caption{Scenario of an UBSer (a) and FBSer (b) perceiving the same graph. The viewer in (a) is a learner of graph theory, who is not informed about the real-world meanings of the graph and perceives the graph from a theoretical perspective. The viewer (b) is a reader of the \emph{Game of Thrones} novel, who knows that the graph presents the relationships of characters in the novel and perceives the graph from an applied perspective. The two viewers have different focus areas of interest while perceiving the graph.}
	\label{fig:figure1}
	\vspace{-0.3cm}  
\end{figure}

%
%
%
%
\IEEEPARstart{A}{} graph is an abstract model that represents relations among entities \cite{A1,A2}. They can often be found in various real- world applications, such as social networks \cite{A73} and computer networks \cite{A74}. In general, graphs can be analyzed from either a theoretical or applied perspective because intrinsic connections exist between abstract graphs and real-world graph data (e.g., high degree nodes in graph theory or influencers in social networks). Such connections are established through background stories that endow entities and relations with real-world meanings and describe the semantics and context of abstract graphs. For example, the graph in \autoref{fig:figure1} shows the relationship network of characters in the \emph{Game of Thrones} novel \cite{A75}. The novel describes a story in which characters from the Sunset Kingdom launched into arduous journeys and built intertwined relationships after the king died in an accident.
\par{Graph perception is an essential part of graph analysis \cite{A76,A70}, in which human viewers explore graph structures of interest and study their meanings through a graph visualization, e.g., a node-link diagram. Does a graph’s background story affect a viewer’s perception? The answer may be yes based on our experience and prior research. The viewers who are unaware of the background story (i.e., UBSers) and the ones who are familiar with the background story (i.e., FBSers) may perform differently when perceiving the same graph. For example, the UBSer in \autoref{fig:figure1}(a) is attracted to the community structures in the graph, whereas the FBSer in \autoref{fig:figure1}(b) is seeking the nodes that represent the protagonists in the background story. Some psychological studies have demonstrated that personal knowledge can affect the focus areas and comprehension of viewers on paintings or artwork \cite{A3,A4,A5}, which motivates our work.
}
\renewcommand{\thefootnote}{\fnsymbol{footnote}}
\footnotetext[1]{Data sets and supporting information: https://github.com/csuvis/Graph-WithStoryData}

\par{Graph perception has been the subject of extensive research \cite{A69,A70,A71}. Some inherent laws of visual perception, such as pre-attentive processing \cite{A8} and the Gestalt laws \cite{A6}, have been used to guide the visual encodings of node-link diagrams to facilitate graph perception. Many aesthetic metrics \cite{A9} and shape constraints \cite{A10} have been proposed for graph layout optimizations to improve the readability of node-link diagrams. Some techniques, such as fisheye interactions \cite{A11} and animations \cite{A12}, utilize the visual context information of target nodes or structures to promote graph perception. Although graph data repositories, such as Pajek \cite{A13} and UCINET \cite{A14}, provide graphs and background stories, limited research has systematically studied the effects of background stories on graph perception.}

\par{To address this gap, this research investigates the effects of background stories on graph perception. We formulated three hypotheses: (H1) background stories can affect the visual focus areas of viewers during interactive graph explorations; (H2) background stories can affect the performance of viewers when identifying specific graph structures, including high degree nodes, bridges, and communities; and (H3) background stories can help viewers construct stable mental models for graph recognition under difficult visual conditions. The three hypotheses covered open-ended (H1) and target-given (H2) perception tasks, as well as visceral (H1), behavioral (H2), and reflective (H3) levels of graph perception \cite{A6,A7}. Twelve real-world graph data sets with accompanying background stories were chosen as the experimental data and were used within three controlled between-subject experiments with 70 participants (i.e., 35 UBSers and 35 FBSers) to test these hypotheses.}

\par{In the experiments, the areas of interest (AOIs) that were interactively selected by the participants, as well as the accuracy and time taken to identify structures and recognize graphs, were recorded. We designed three similarity indicators to quantitatively measure the differences of AOI selections between the FBSers and UBSers. We also conducted a series of significance analyses to understand the significant differences between the performances of the FBSers and UBSers in structure identification and graph recognition. In addition, we collected the opinions and feedback of the participants for qualitative analysis. The results found that background stories: 1) affected the visual focus areas of the participants, 2) significantly affected the performance of the participants when identifying community structures but not high degree and bridge structures, and 3) influenced graph recognition under difficult visual conditions.}
\par{In summary, this research presents the first attempt to evaluate the effects of background stories on graph perception. It contributes three experimental methodologies, a series of instructive findings, and a set of prepared graph data with accompanying background stories. This research provides new insights into the inherent laws of graph perception and brings new considerations into the design decisions of storytelling visualizations and interactive graph explorations. In addition, this research should inspire researchers to investigate the effects of background stories or other personal knowledge on the visual perception of abstract data models, such as trees, tabulations, or trajectories.}

\section{Related Work}
\subsection{Graph Visualization}
Graph visualization has long been an active research topic \cite{A15,A16,A17,A95}. Node-link diagrams are a popular graph visualization method due to their prominent superiority in visual perception and interactive analysis \cite{A18}. However, their main limitation is that visual clutter often occurs when visualizing a massive graph \cite{A19}. Many methods have been proposed to address this issue. Graph reduction methods, such as clustering \cite{A20}, filtering \cite{A21}, and sampling \cite{A22,A90,A92,A93}, reduce the size of a graph for the easy perception of important structures. Advanced layout \cite{A23,A24,A94} and edge-bundling algorithms \cite{A25,A26} optimize the spatial arrangement of nodes and edges to reduce visual clutter. Interactions, such as zooming and panning \cite{A27}, help users observe local details. In addition to these methods, the inherent laws of graph perception have been studied to provide a guideline for node-link diagram design. Our work belongs to this category.

\subsection{Graph Perception}
\label{sec:section2.2}
There are three perception levels that existing graph perception studies focus on, i.e., visceral, behavioral, and reflective. These levels are based on the work of Norman \cite{A28} and the recommendation of Bennett \cite{A6}.
\par{The visceral level of graph perception is one where human viewers form first impressions about graph visualizations and then identify prominent visual features by intuition \cite{A6}. Two important perception theories, namely, pre-attentive rules \cite{A8} and Gestalt laws \cite{A6}, work at this level. Ware et al. \cite{A8} demonstrated that using the pre-attentive rules to encode graph features can help viewers quickly distinguish these features in node-link diagrams. Marriott et al. \cite{A7} confirmed that if graph layouts take full account of symmetry and continuity rules in the Gestalt laws, then viewers can form deep impressions on graph structures.}
\par{The behavioral level of graph perception is related to the identification of meaningful graph structures \cite{A6}. The readability of graph layouts has been found to have a strong impact on structure identification. Many measurable metrics have thus been proposed to evaluate readability. For example, Purchase et al. \cite{A9} proposed the metrics of edge crossings and edge bends to measure the  aesthetic readability of graph layouts. Dunne et al. \cite{A29} proposed the node occlusion and group overlapping metrics to identify problematic visual shapes. Taylor et al. \cite{A30} designed the homogeneity and concentration metrics to measure the usage of display areas in node-link diagrams.}
\par{The reflective level emphasizes the understandability of visualizations \cite{A6}. To achieve this goal, viewers are required to obtain and utilize context information sufficiently. Psychological researchers \cite{A31} have divided contexts into displayed and non-displayed ones. Displayed contexts are visual features that are presented on the screen but are not target features. Previous studies have found that displayed contexts can help viewers understand target structures in node-link diagrams. For example, Fisheye interactions magnify displayed contexts around target structures for graph perception \cite{A11}. Gorochowsk et al. \cite{A12} found that relatively stable substructures at certain spatial positions can help viewers perceive graph changes. However, to the best of our knowledge, limited work has been conducted to systematically investigate the effects of non-displayed contexts on graph perception. The background story of a graph is a non-displayed context.}
\subsection{Visual Perception with Non-displayed Context}
Non-displayed contexts refer to the invisible knowledge and experience of viewers \cite{A3,A7}. Some psychological studies have shown that knowledge and experience can affect the visual perception of pictures and artwork. In terms of knowledge, Rahman et al. \cite{A3} found that knowledge can shape perception by penetrating early visual processes. Lupyan \cite{A32} proved that personal knowledge can enrich the visual perception of paintings or pictures. Humphrey et al. \cite{A4} discovered that domain knowledge can moderate the influence of visual saliency in scene recognition. In terms of experience, Todorovic \cite{A33} found that visual elements tended to be grouped if they had been perceived together during the past experiences of viewers. Braly \cite{A34} discovered that the past experience of viewing certain visual forms can influence the subsequent perception of other visual forms. Wiley \cite{A5} demonstrated that an individual’s artistic experience can affect visual perception styles. These studies inspired the present work. Background stories of graphs could be considered to be a type of personal knowledge and node-link diagrams could be considered to be stimuli (similar to paintings and pictures). Therefore, the present work seeks to understand the degree to which these non-displayed contexts influence graph perception.

\section{Hypotheses}
The background story of a graph refers to the textual description that introduces related information about the graph, such as the real-world meanings of nodes and edges, the semantics, and the context of the graph. This study aims to conduct controlled experiments to investigate the effects of background stories on graph perception. To guide the experimental design, the following hypotheses were formulated:
\par{\textbf{H1:} Background stories can affect the visual focus areas of viewers during open-ended graph explorations. A number of previous studies in psychology have demonstrated that the focus areas of viewers while looking at pictures can be affected by personal knowledge \cite{A4}. From an analogy point of view, node-link diagrams are visual stimuli similar to pictures and background stories can be regarded as a type of personal knowledge. Thus, we propose that the focus areas of UBSers, who are unaware of background stories, will be different from those of FBSers, who are familiar with background stories. \autoref{fig:figure1} shows an example.}
\par{\textbf{H2:} Background stories can affect the performance of identifying graph structures. Previous studies have found that the three types of graph structures (e.g., high degree node, bridge, and community) can be widely perceived in node-link diagrams \cite{A22,A35}. Background stories may indicate the existence of the three types of graph structures. For example, a story mentioned that the Sharpstone Auto (SA) baseball team had three popular players, which indicated the existence and even number of high degree nodes in the friendship network of the team \cite{A36}. Thus, we suppose that FBSers perform better than UBSers when identifying the three types of graph structures.}
\par{\textbf{H3:} Background stories can help viewers construct stable mental models of graphs. The mental models refer to memorable and recognizable visual patterns that viewers observe from node-link diagrams and form in their mind \cite{A37,A38,A39}. Background stories may indicate the existence of stable visual patterns, such as unique overall shapes and anti-interference visual features. For example, a story described that a two-part fission event occurred in the Zachary’s karate club \cite{A40}, which indicated that the member friendship network may have a symmetrical shape formed by two warring communities. Thus, we believe that the mental models of FBSers are more stable than those of UBSers. Stable mental models should enable FBSers to perform better than UBSers when recognizing graphs that they have seen previously.}

\section{Experimental Design}
Three experiments were designed to evaluate the three hypotheses. The first experiment (Ex1) verified whether background stories can affect the focus areas of participants during open-ended graph explorations (H1). The second experiment (Ex2) examined whether background stories can affect the performance of identifying high degree, bridge, and community structures (H2). The third experiment (Ex3) investigated whether background stories can help participants construct stable graph mental models for graph recognition (H3). In this section, the design of the experiments is detailed.

\begin{figure*}[htb]
	\centering
	\setlength{\abovecaptionskip}{0.1cm}   
	\setlength{\belowcaptionskip}{-0.5cm}   
	\includegraphics[width=0.97\textwidth]{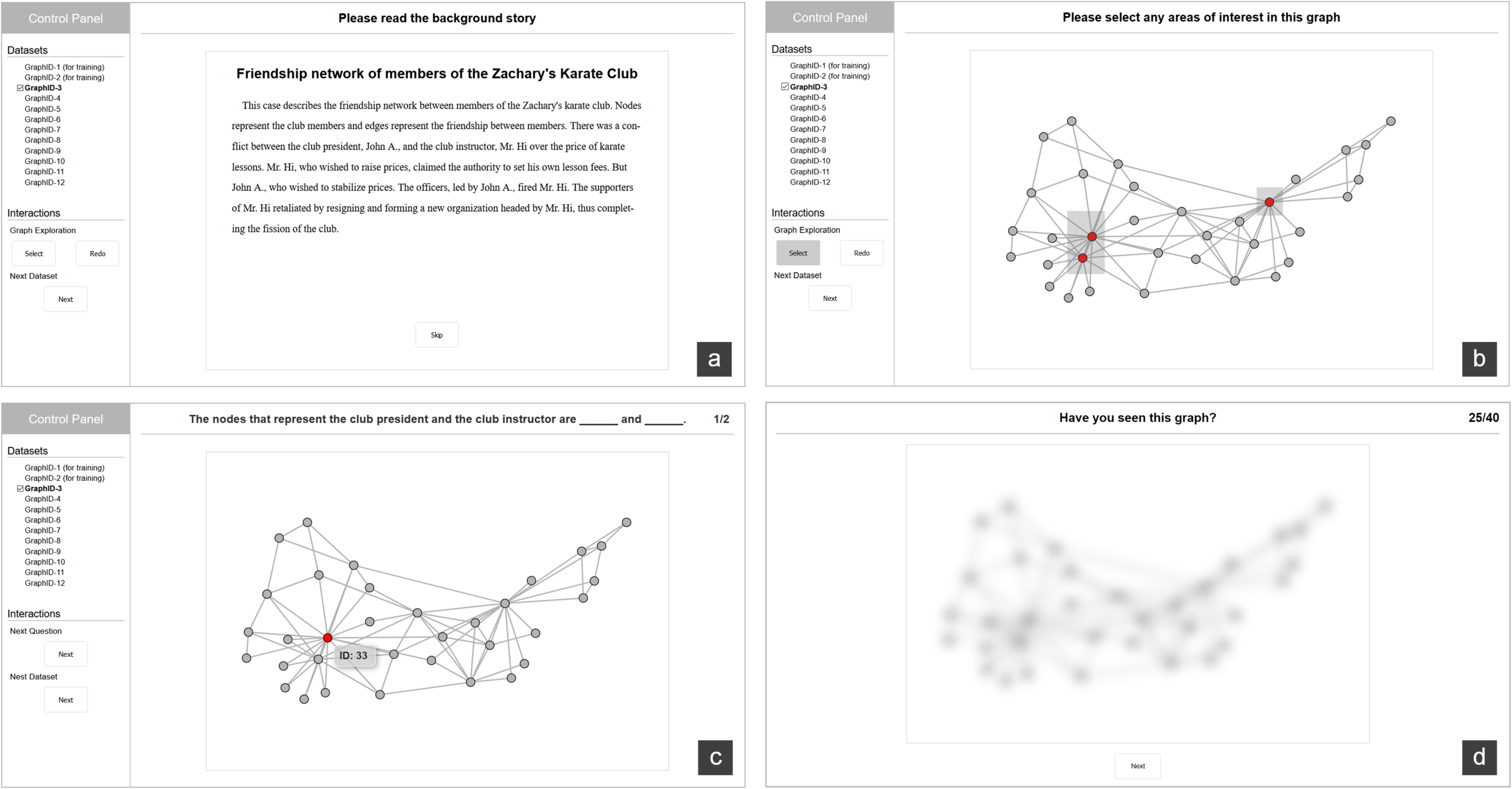}
	\caption{Interfaces used by the FBSers in the three experiments. (a) The background story of the member friendship network of the Zachary’s karate club. (b) The visualization of the network used in Ex1, where a FBSer has selected two areas of interest. (c) The visualization result of the network in Ex2, where a FBSer has selected the node of ID-33 as one of the answers to the current objective question. (d) A blurred graph used in Ex3, where participants answered if they had seen this graph before.}
	\label{fig:figure2}
\end{figure*}
\begin{table*}[!ht]
	\scriptsize
	\renewcommand\tabcolsep{5.1pt}  
	\renewcommand\arraystretch{1.5} 
	\vspace{0.1cm}
	\centering
	\caption{Basic information on the experimental graphs and underlying structures for each graph in Ex2. The number of nodes and edges were obtained using the TULIP software \cite{A41}.}
	\label{tab:table1}
	\begin{tabular}{cllcccccccc}
		\toprule  
		\multirow{2}{*}{$\bf{Graph }$ $\bf{ ID}$}&\multicolumn{1}{c}{\multirow{2}{*}{$\bf{Graph} $ $\bf{Name}$}} &\multicolumn{1}{c}{\multirow{2}{*}{$\bf{Graph} $ $\bf{Type}$}}
		&\multirow{2}{*}{$\bf{Nodes}$}& \multirow{2}{*}{$\bf{Edges}$}& \multicolumn{3}{c}{\multirow{1}{*}{$\bf{Inquired}$ $\bf{Structure}$ $\bf{in}$ $\bf{Ex2}$}}&\\
		\cline{6-8}
		\multicolumn{1}{c}{}&\multicolumn{1}{c}{}&\multicolumn{1}{c}{}&\multicolumn{1}{c}{}&\multicolumn{1}{c}{}&$\bf{High} $ $\bf{Degree}$&$\bf{Bridge}$&$\bf{Community}$&\\
		
		\midrule  
		\hypertarget{GD1}{GD1}&Co-appearance network of characters in \emph{Les Miserables} (for training)&Co-appearance Network&77&254&Q1&&\\	\hypertarget{GD2}{GD2}&Animal social network of bottlenose dolphins (for training)&Animal Network&62&159&&Q3&Q2\\	\hypertarget{GD3}{GD3}&Friendship network of members of the Zachary's karate club&Friendship Network&34&78&Q4&&Q5\\	\hypertarget{GD4}{GD4}&Relationship network of characters in \emph{Game of Thrones}&Relationship Network&107&352&&&Q6\\	\hypertarget{GD5}{GD5}&Friendship network of the Transatlantic Industries (TI) baseball team&Friendship Network&13&37&Q7&& \\	\hypertarget{GD6}{GD6}&Games schedule network of American college football&Sport Network&115&615&&&Q8\\	\hypertarget{GD7}{GD7}&Social network of employees in a wood-processing facility&Social Network&24&38&&Q10&Q9\\	\hypertarget{GD8}{GD8}&Network of states and legal bases for divorce&Affiliation Network&59&225&Q11&&\\	\hypertarget{GD9}{GD9}&Co-purchasing network of political books in the United States&Co-purchasing Network&105&441&&&Q12\\	\hypertarget{GD10}{GD10}&Friendship network in a German boys’ school class&Friendship Network&48&179&Q13&&\\	\hypertarget{GD11}{GD11}&Strong political tie network in a midwestern county in the US&Political Tie Network&14&56&&Q14& \\	\hypertarget{GD12}{GD12}&Friendship network of the Sharpstone Auto (SA) baseball team&Friendship Network&13&39&Q15&&\\	
		
		\bottomrule
	\end{tabular}
\end{table*}

\subsection{Graph Data Sets}
We initially gathered 38 popular real-world graph data sets as candidates. Then, we selected 12 graph data sets from the candidates for the three experiments, as shown in \autoref{tab:table1}. The following criteria were used when selecting the 12 data sets: (1) Graphs should cover a range of topics, e.g., social, friendship, sport, animal, and affiliation networks. The meanings of their nodes and edges should be familiar and easily understood so that comprehension difficulties can be reduced and participants’ interest can be stimulated. (2) Small graphs were prioritized because large graphs would result in poor readability and overburden participants’ visual perception. (3) Graphs needed to have rich structures to attract participants attention during Ex1, contain at least one of the three types of structures for Ex2, and should present distinct overall shapes and visual features for Ex3. (4) Graphs should have interesting background stories that can be edited or condensed to fit within the duration of the experiments. Moreover, all graphs were treated as undirected graphs.

\par{A number of steps were undertaken to process the background stories of the 12 graphs. First, we carefully reviewed the literature and previous studies to collect multiple story versions for each graph. Second, pilot studies were run and found that background stories which were 70 to 120 words enabled participants to finish reading within a short time. Third, we shortened and edited the background stories, along with the objective question design in Ex2, ensuring that key information was retained and trivial information was discarded. The background stories of the 12 graphs are provided in the supplementary materials of this paper.}

\subsection{Participants}
Seventy participants were recruited to participate in the experiments (i.e., 35 males and 35 females). All participants had a normal or corrected-to-normal vision. Their ages ranged from 20 to 31 years old (with a mean value of 24). They were all undergraduate or graduate students. Forty of the participants had a computer science background, and the other 30 had non-computer science backgrounds. The participants were randomly divided into two groups: FBSers and UBSers. Each group consisted of 20 participants with a computer science background and 15 participants with non-computer science backgrounds. The FBSers were asked to learn the background stories of the graphs in the experiments, whereas the UBSers were not allowed to read the background stories during the experiments. Before grouping, we asked each participant if they were familiar with the provided background stories to avoid assigning such participants to be UBSers.

\subsection{Visual Stimuli}
The visual stimuli in the three experiments were the node-link diagrams that shared visual encodings and graph drawing settings. As shown in \autoref{fig:figure2}, a node-link diagram was drawn within a rectangle area of 900 × 620 pixels using a white background. A gray circle with a radius of about 8.43 pixels represented a node, a gray line with a width of about 2.5 pixels represented an edge, and the graph layout was generated by the Fast Multipole Embedder algorithm \cite{A41,A42,A81,A82}. The layout of the same graph was consistent across all three experiments. These encodings and settings were derived from pilot studies because they generated relatively high readability measured by the metrics of edge crossings, edge crossings angle, angular resolution (deviation) \cite{A83,A84,A29} and density \cite{A81}. The measurement results are provided in the supplementary materials.

\subsection{Tasks}
\subsubsection{Experiment 1}
Ex1 aimed to understand how background stories influenced the focus areas of viewers during open-ended graph explorations. The participants were asked to select any AOIs without limiting the number of  selections in the node-link diagrams generated by the 12 graphs, as shown in \autoref{fig:figure2}(b). The nodes or subgraphs selected by the participants were representations of their focus areas when observing the node-link diagrams. The selection results would help verify H1.

\subsubsection{Experiment 2}
Ex2 was designed to examine whether background stories can affect the performance of identifying the three types of graph structures, i.e., high degree, bridge, and community structures. The participants were asked to answer predetermined objective questions that probed a specific case of the three types of graph structures in the 12 graphs.
\par{The objective question design had two key elements: (1) As we needed to control the number and difficulty of questions to ensure that the participants could finish the questions within a limited time, we designed no more than two objective questions for each graph and 15 objective questions in total (\autoref{tab:table1}). Thus, the graph structures with relatively high visual saliency were selected and the three structure types were balanced across the selection. (2) Each question required the preparation of two descriptive versions that were tailored to the FBSers and the UBSers but probed the same graph structure. To design the FBSer version, words that were in the background story and related to the probed structure were used. General graph structure descriptions were avoided. Conversely, only general structure descriptions were used to design the UBSer version. Several pilot studies were conducted to refine the two descriptive versions of each question to ensure that they had a moderate difficulty and that there was consistency within the probed structure. All objective questions are provided in the supplementary materials.}
\par{Taking the member friendship network of the Zachary’s karate club as an example, the background story is that the club president and an important club instructor had a serious conflict over the price of karate lessons. The supporters of the instructor gradually formed a new organization, thereby causing fission within the club. In accordance with the story and the feedback gathered in pilot studies, two objective questions were designed. The first question for the FBSers was \emph{which nodes represent the president and instructor?} The first question for the UBSers was \emph{which nodes are the top two highest degree nodes in this graph?} For the second question, the FBSers were asked \emph{how many factions are in the club?}, whereas the UBSers were asked \emph{how many communities exist in this graph}?}
\par{Two subjective questions were set to collect information on the feelings of participants on the difficulty and confidence levels of answering each objective question. The two subjective questions were rated by using a five-point Likert scale, ranging from 1 (lowest level) to 5 (highest level).}

\subsubsection{Experiment 3}
Ex3 investigated whether background stories can help viewers construct stable mental models for graph recognition. The participants were asked to complete a set of graph recognition trials. In each trial, either one of the 12 graphs or an interference graph was presented and the participants were required to judge whether they had previously seen the graph, as shown in \autoref{fig:figure2}(d).
\par{Generating moderately difficult recognition conditions was crucial because existing psychological research has shown that personal knowledge has a measurable effect on perceptual analysis under moderately difficult conditions \cite{A3,A34,A43}. Thus, interference graphs and blurred visualizations were used in Ex3.}
\begin{figure}[!t]
	\centering
	\setlength{\abovecaptionskip}{0.1cm}   
	\includegraphics[width=0.95\columnwidth]{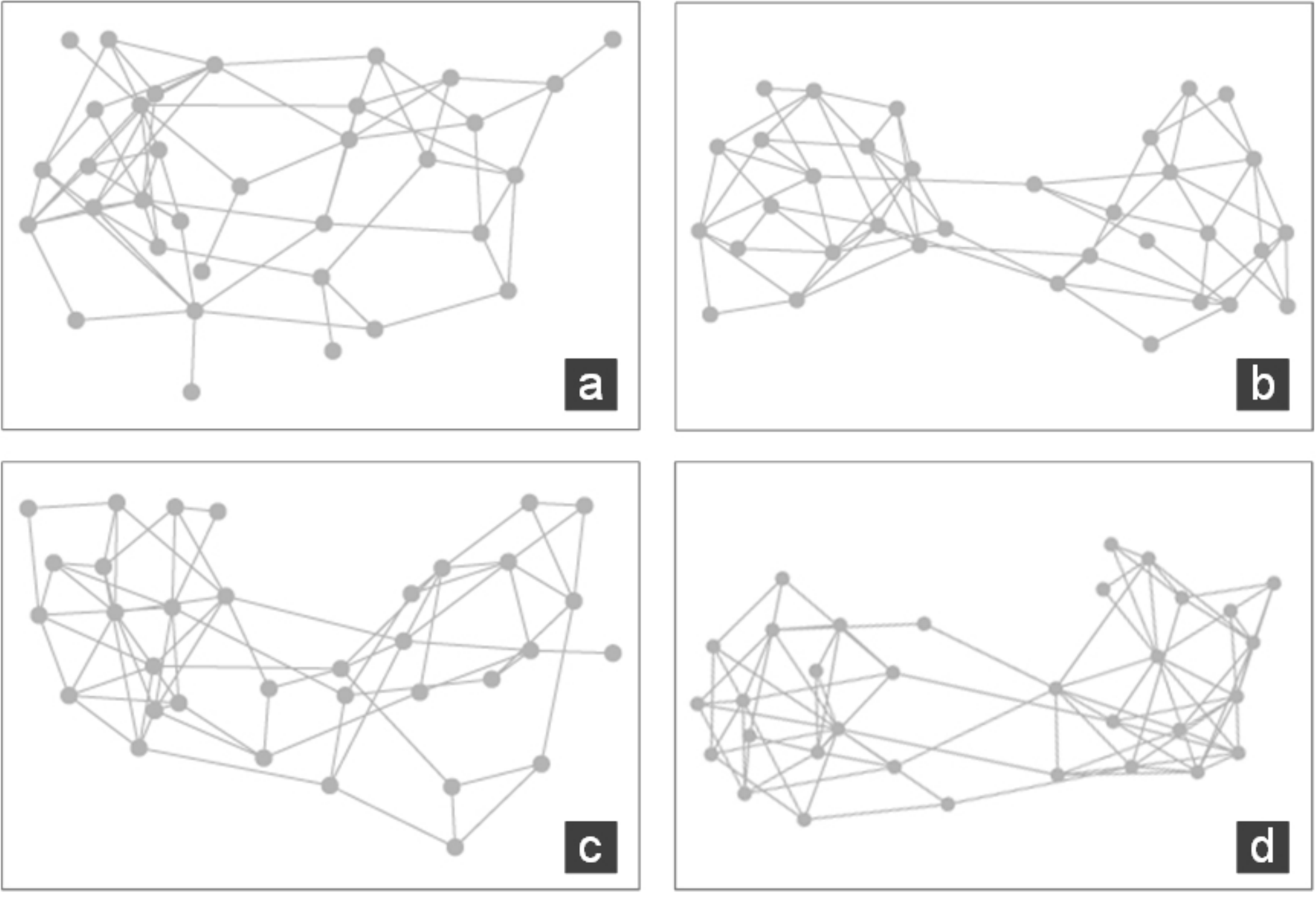}
	\caption{Four generated interference graph candidates of the member friendship network of the Zachary’s karate club (\autoref{fig:figure2}). Candidates (a) and (d) had a low and high perceptual similarity with the raw graph, respectively, whereas candidates (b) and (c) had a moderate perceptual similarity, according to the feedback gathered in pilot studies.}
	\label{fig:figure3}
\end{figure}
\begin{figure}[!t]
	\centering
	\setlength{\abovecaptionskip}{0.1cm}   
	\includegraphics[width=0.95\columnwidth]{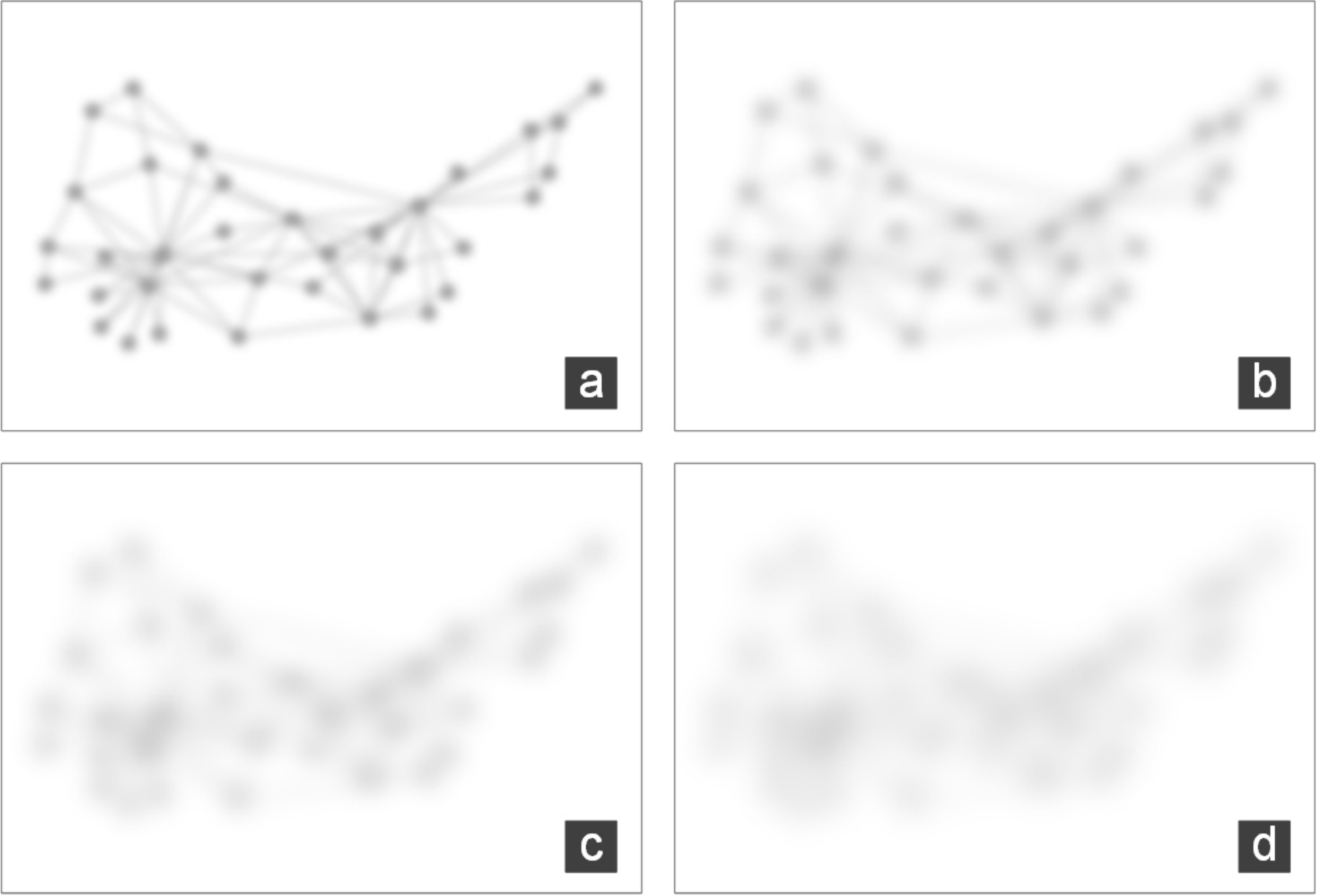}
	\caption{ Four blurred node-link diagrams with blurring radii of (a) 2.5, (b) 4.5, (c) 6.5, and (d) 9.5, respectively. The blurred diagrams (a) and (b) had high and moderate recognizability, whereas diagrams (c) and (d) had low recognizability, according to the feedback gathered in pilot studies.}
	\label{fig:figure4}
	\vspace{-0.3cm}
\end{figure}
\par{The interference graphs used in Ex3 were graphs that were not the 12 raw graphs but were similar to them. Selecting or generating interference graphs with a moderate difficulty was challenging because the interference graphs and raw graphs needed to be perceptually similar in an overall sense so that they could be plausibly grouped into one category, but also adequately distinct in details to be assigned to different categories \cite{A43}. Thus, a four-step process that integrated the advantages of automatic generation and manual selection was used to obtain appropriate interference graphs. (1) We stipulated that an interference graph only needed to be similar to one raw graph. This controlled the difficulty of graph perception and interference graph generation. (2) We estimated the structural properties of each raw graph, including the number and sizes of communities and the probability of edges within and between communities. (3) We used a random partition generator within NetworkX \cite{A44} to randomly generate at least 50 interference graph candidates for each raw graph. The generator guaranteed that each interference graph candidate was similar to the reference raw graph in terms of structural properties. However, the perceptual similarity was unclear. (4) We invited 10 volunteers to visually compare each candidate with the reference raw graph and rate a perceptual similarity score to each candidate by a seven-point Likert scale, ranging from 1 (lowest similarity) to 7 (highest similarity). The candidates with a mean rating of 4 were then assumed to be appropriate interference graphs with a moderate similarity (difficulty), as shown in \autoref{fig:figure3}.

\par{The difficult visual conditions used in Ex3 included mirroring and blurring, which have been widely used in perception studies \cite{A45,A46,A79}. For mirroring, horizontal flip to mirror node-link diagrams were used because they presented a moderate difficulty during pilot studies, whereas the vertical flip was difficult. For blurring, Gaussian blur was used to blur node-link diagrams \cite{A47}. To determine the moderate blurring radius, ten radii were initially selected (i.e., 1.5, 2.5, 3.5, 4.5, 5.5, 6.5, 7.5, 8.5, 9.5, and 10.5). If the radius of Gaussian blur was large, edges and nodes in node-link diagrams became difficult to identify clearly (Figure 4). Then, ten volunteers were asked to compare each raw graph and blurred raw graphs with different radii and rate a subjective recognizability score of each blurred raw graph on a seven-point Likert scale, ranging from 1 (lowest recognizability) to 7 (highest recognizability). The recognizability probed whether viewers could distinguish the raw graph from the blurred raw graphs. Finally, all blurred raw graphs with a score of 4 were chosen and it was found that the mean of their blurred radii was 4.5. Therefore, a blurring radius of 4.5 was used because this setting can generate moderate recognizability, as shown in \autoref{fig:figure4}.

\par{There were two graph-displaying modes considered: display multiple graphs at once or display one graph at a time. We opted to display one graph at a time because it would increase participants' concentration and would reduce any perceptual bias caused by size-reduced graphs and comparable hints that could be found when comparing multiple graphs.}
\par{To determine the number of trials to use, we looked towards the data from pilot studies. This data determined that the volunteers began losing their focus after viewing 50 graphs due to fatigue. Thus, forty trials was chosen as the number of trials to use in Ex3. The 40 trials included 10 blurred raw graphs, 10 blurred and flipped raw graphs, and 20 blurred interference graphs, where each raw graph had two interference graphs. The ratio of raw graphs to interference graphs was 1 to 1, which was consistent with the rule of moderate difficulty. Moreover, eight trials were also created for the two training graphs.
}
\par{Furthermore, two subjective questions were set to collect the subjective opinions of participants about the difficulty of each trail and the confidence level while doing so. The two subjective questions used a five-point Likert scale, ranging from 1 (lowest level) to 5 (highest level).}

\subsection{Interface and Apparatus}
An interactive visualization system was developed to support the experimental tasks. As shown in \autoref{fig:figure2}, the left section of the interface contained the control panel, which listed the experimental graphs, highlighted the currently examined graph, and provided functional buttons. The right section of the interface had a node-link diagram and a tip area showing the current task. Three lightweight interactions were provided to help the participants perform tasks in Ex1 and Ex2. (1) A range-selecting interaction enabled the participants to interactively select nodes or areas by using the mouse to draw rects in Ex1. (2) A redo interaction allowed the participants to cancel the latest selection in Ex1. (3) A hovering interaction helped the participants obtain the ID of a certain node by hovering over the node with the mouse in Ex2. All experiments were conducted on a desktop computer with a 3.4 GHz processor and 16 GB of RAM, and a 24-inch monitor with a 1920 × 1200 resolution. A standard wired mouse and wired keyboard were used.

\subsection{Procedures}
All participants took part in all three experiments. The FBSers were required to read the background stories before performing the tasks, whereas the UBSers were not allowed to read the background stories throughout the experiments. Using this between-subject design avoided the influence of learning effects with the UBSers, that is, acquiring related information introduced in the stories, by drawing lessons from previous perception studies  \cite{A3,A85,A86}. Objective and subjective questionnaires were used to collect the answers of the participants. Examples of the questionnaires are provided in the supplementary materials.

\subsubsection{Experiment 1}
The procedure of Ex1 consisted of four stages: introduction, tutorial, training, and formal study. The task of Ex1 was to select AOIs in each of the 12 graphs. The  \hyperlink{GD1}{GD1} and \hyperlink{GD2}{GD2} were for training, and the other 10 graphs were used within the formal study.
\par{Introduction stage: The participants filled out a demographic survey with information about age, gender, and educational background. The instructors introduced the purpose of the study, the tasks, and the procedures of the three experiments and demonstrated the interface and questionnaires. The instructors also introduced basic concepts related to abstract graph models, node-link diagrams, typical graph structures, and real-world meanings of abstract models. This stage took approximately 15 min.}
\par{Tutorial stage: This stage familiarized the participants with the interface and task. The instructors guided the participants in completing the first training graph. This stage took approximately 5 min.}
\par{Training stage: This stage ensured that the participants mastered how to use the interface to complete the task. The participants were asked to complete the task using the first and second training graphs. They were allowed to ask questions. If they were confused with the interface or task, the participants would be reverted back to the previous stage. This stage took approximately 10 min.}
\par{Formal study stage: The participants were required to complete the AOI selection task on the 10 graphs one by one (i.e., 10 trials). The graph order was randomized to mitigate learning effects. Before performing the task on a graph, the FBSers were given 1 min to read the background story of the graph. Assuming a normal reading speed and story length of 120 words, this duration enabled the FBSers to read the story up to three times if desired. This step was not preformed by the UBSers. After completing all graphs, we conducted a short interview. The participants were allowed to review their selection results and encouraged to provide their thoughts. This stage had no time limit. Most of the participants were able to complete a trial within 2 min. Thus, this stage took approximately 30 min.}

\subsubsection{Experiment 2}
After a 10 min break, the participants proceeded to Ex2. The task of Ex2 was to answer predetermined objective questions in each graph. The procedure of Ex2 was similar to that of Ex1. In the tutorial and training stages, the instructors helped the participants familiarize themselves with the task by using the two training graphs. In the formal study stage, the participants needed to complete the task on the 10 graphs one by one with a random order (i.e., 10 trials), and the FBSers were allowed to read the background stories before performing the task. After completing an objective question, the participants needed to rate the difficulty and confidence levels of answering each question on the basis of their subjective feelings using a five-point Likert scale. After completing all graphs, a short interview was conducted, and the participants reviewed their answers and provided their feedback. No time limit was imposed during Ex2. The participants generally completed a graph within 3 min. Thus, the formal study stage took approximately 40 min.

\subsubsection{Procedure of Experiment 3}
After Ex2, the participants rested for at least 15 min, and then proceeded to Ex3. The task of Ex3 was to answer whether a graph visualized in a blurred node-link diagram had been seen before or not. The tutorial and training stages of Ex3 were similar to those of Ex1 and Ex2. The formal study stage consisted of learning and testing sessions. In the learning session, the participants were asked to view all 10 raw graphs one by one. The display time of a raw graph was 1 min. During this time, the background story of a raw graph was displayed alongside the raw graph for the FBSers. If the participants were convinced that they had memorized the graph firmly, they could proceed to the next graph. In the testing session, the participants were asked to complete 40 trials. Each trial displayed a single blurred node-link diagram that was possibly generated by a raw graph, a flipped raw graph, or an interference graph. The participants answered whether they had seen the graph before or not without time limits. The raw graphs and flipped raw graphs were considered as the graphs that had been seen previously. Similar to Ex2, after completing a trial, the participants needed to rate the difficulty and confidence levels of the trial by using a five-point Likert scale. After completing all trials, a short interview was conducted. The formal study stage took approximately 40 min. The participants who completed the three experiments were compensated with \$10 per hour.

\section{Experimental Results}
\subsection{Analysis Approach}
The experimental results can be divided into quantitative and qualitative parts. Quantitative results included the AOI selections and selection sequences of each participant in each graph (Ex1) and the accuracy and time of each participant in completing each objective question (Ex2) or trial (Ex3). Qualitative results were the subjective feelings of each participant in answering each objective question (Ex2) or trial (Ex3) in terms of difficulty and confidence levels and the feedback of the participants in the interviews (Ex1, Ex2, and Ex3). In addition to mean values and standard deviation calculations, the results were analyzed from two main aspects as follows.
\par{\textbf{Similarity measurement:} This analysis approach was for Ex1. We designed three similarity indicators to quantitatively measure the differences of focus areas between the FBSers and UBSers in a graph because the results of Ex1 were unfit for traditional significance analysis. The value ranges of the three indicators were from 0 to 1, with larger values indicating a higher degree of similarity. The definitions and calculations of the indicators are detailed as follows.}

\par{(1) The \emph{spatial-distribution similarity} (PS) of AOIs measured the similarity between two AOI distributions in a graph layout regardless of the structural meaning of AOIs. The two distributions were generated from the AOIs selected by the FBSers and UBSers in the same graph, respectively. The calculation of PS had four steps. First, we regarded the layout of a graph as an \textit{n × m} pixel-based position matrix notated as\[
	PM_{t,k}=\begin{bmatrix}
	p_{11} & \dots & p_{1j} & \dots & p_{1m} \\
	\dots & \dots & \dots & \dots & \dots \\
	p_{i1} & \dots & p_{ij} & \dots & p_{im} \\
	\dots & \dots & \dots & \dots & \dots \\
	p_{n1} & \dots & p_{nj} & \dots & p_{nm}
	\vspace{1ex} 	
	\end{bmatrix}
	\]
	where ${t}$ and ${k}$ represent the type of participants and the graph ID, respectively; ${p_{ij}}$ represents the number of entries of the pixel at row ${i}$ and column ${j}$ with an initial of 0. Second, we constructed a mapping table to convert each node in the layout to the corresponding pixels in the matrix. Third, we counted the entries made by the FBSers and UBSers for each node and obtained ${\textbf{\emph{PM}}_{FBSers,k}}$ and ${\textbf{\emph{PM}}_{UBSers,k}}$ based on the mapping table. Finally, we calculated the similarity of the two matrices through the computation of a min-max normalization of Euclidean distance \cite{A48}.}
\par{(2) The \emph{structure similarity} (SS) of AOIs measured the similarity between the specific graph structures concerned by the FBSers and UBSers in the same graph. We selected four representative types of graph structures, namely, high degree, bridge, community, and other structures, because the former three types are popular in graph perception \cite{A22,A35} and were examined in Ex2. The calculation of SS involved three steps. First, we manually categorized all AOIs into the four types. Second, we counted the entries made by the FBSers and UBSers for each type and constructed two structure vectors notated as:
	\[
	\setlength\abovedisplayskip{0.2cm}
	\setlength\belowdisplayskip{0.2cm}
	SV_{t,k}=\left ( {{s_h},{s_b},{s_c},{s_o}} \right) \]where ${s_{h}}$, ${s_{b}}$, ${s_{c}}$, and ${s_{o}}$ represent the numbers of high degree, bridge, community, and other structures, respectively. Third, we calculated the similarity between ${\textbf{\emph{SV}}_{FBSers,k}}$ and ${\textbf{\emph{SV}}_{UBSers,k}}$ by using a min–max normalization Euclidean distance.}
\par{(3) The \emph{order similarity} (OS) measured the similarity between the selection orders of different types of structures performed by the FBSers and UBSers for the same graph. Its calculation was based on the four structure types and the selection sequences of the participants. First, we counted the entries of the FBSers and UBSers for each type in order of 1st, 2nd, 3rd, and others in a graph. Then, we constructed two matrices for the FBSers and UBSers, respectively, notated as\[
	OM_{t,k}=\begin{bmatrix}
	o_{1-h} & o_{1-b} & o_{1-c} & o_{1-o} \\
	o_{2-h} & o_{2-b} & o_{2-c} & o_{2-o} \\
	o_{3-h} & o_{3-b} & o_{3-c} & o_{3-o} \\
	o_{o-h} & o_{o-b} & o_{o-c} & o_{o-o}
	\end{bmatrix}
	\]
where the four rows represent the orders of 1st, 2nd, 3rd, and others, respectively; the four columns represent the four types of graph structures, respectively; an element represents the number of entries of a certain structure type in a specific order. Finally, we calculated the similarity between ${\textbf{\emph{OM}}_{FBSers,k}}$ and ${\textbf{\emph{OM}}_{UBSers,k}}$ by using min–max normalization Euclidean distance.}
\begin{figure*}[!ht]
	\centering
	\vspace{-0.03cm}  
	\setlength{\abovecaptionskip}{0.1cm}   
	\setlength{\belowcaptionskip}{-0.6cm}
	\includegraphics[width=18cm]{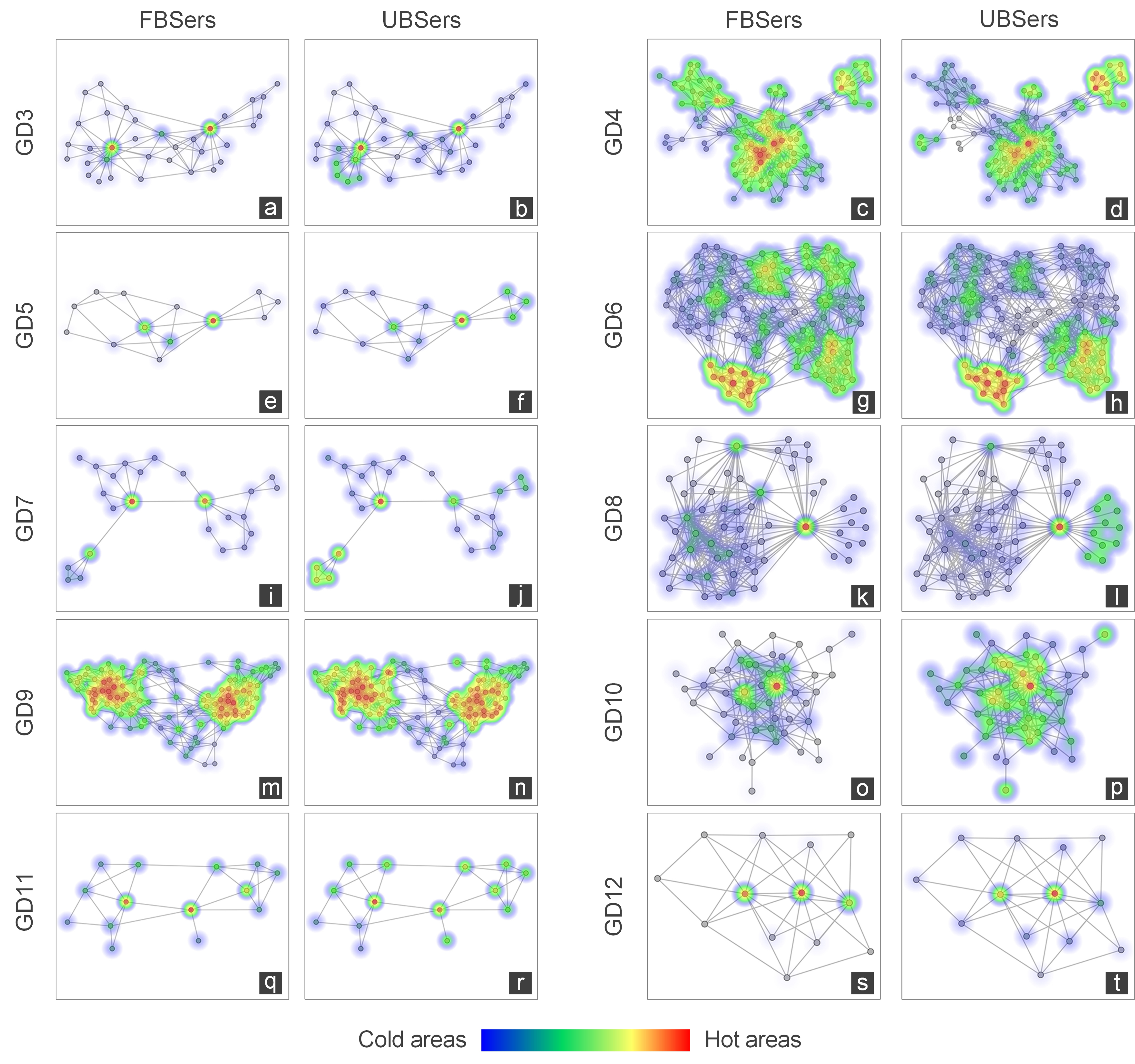}
	\caption{Results of AOI selections made by the FBSers and UBSers per graph in Ex1. Nodes or areas that have a large number of entries are represented by warm colors, whereas nodes or areas that have a small number of entries are represented by cold colors.}
	\label{fig:figure5}
\end{figure*}
\par{\textbf{Significance analysis:} This analysis approach was for Ex2 and Ex3. A series of significant tests were performed to examine the accuracy differences between the FBSers and UBSers in Ex2 and Ex3. We used Shapiro–Wilk tests to examine the normality of experimental results. As most results did not follow the normal distribution $\left ( \rho < 0.05 \right )$, we thus used non-parametric Kruskal–Wallis tests to examine significant differences $\left ( \rho < 0.05 \right )$. We also reported Cohen's d to communicate effect sizes \cite{A89}. If a significant difference was found, then the winner was determined based on the corresponding mean values of accuracy.}

\subsection{Results of Experiment 1}	
We assumed that background stories can affect the focus areas of the participants during open-ended graph explorations (H1). The mean values of the three similarity indicators of the 10 graphs (OS: $\mu$ = 0.46; SS: $\mu$ = 0.47; PS: $\mu$ = 0.58) fell into the medium similarity level (i.e., 0.4, 0.6), \textbf{thereby indicating that the focus areas of the FBSers were moderately similar to those of the UBSers.} That is, the background stories affected the graph explorations of the FBSers. We analyzed the indicator results shown in \autoref{tab:table2} and the \textbf{\emph{PM}} results visualized by using the heatmaps with the same scale \cite{A80} (\autoref{fig:figure5}) as follows. Detailed \textbf{\emph{SV}} and \textbf{\emph{OM}} results are provided in the supplementary materials.

\begin{table}[!t]
	\scriptsize
	\renewcommand\tabcolsep{20pt}  
	\renewcommand\arraystretch{1.5} 
	\vspace{0.15cm}
	\centering
	\caption{Quantitative results of Ex1 in terms of the three similarity indicators. We define three similarity levels as low [0, 0.4], medium (0.4, 0.6), and high [0.6, 1] and denote them as “L”, “M”, and “H”, for reading convenience. }
	\label{tab:table2}
	
	\begin{tabular}{clll}
		
		\hline  
		
		\multirow{2}{*}{\textbf{Graph ID}} &\multicolumn{3}{c}{\textbf{Similarity Indicators}}\\
		\cline{2-4}

				&\multicolumn{1}{c}{\textbf{OS}}& \multicolumn{1}{c}{\textbf{SS}}& \multicolumn{1}{c}{\textbf{PS}}\\ 		
		\hline
		\multicolumn{1}{c}{\hyperlink{GD3}{GD3}}&0.44 M&0.42 M&0.61 H\\	
		\hline
		\multicolumn{1}{c}{\hyperlink{GD4}{GD4}}&0.57 M&0.60 H&0.46 M\\				
		\hline
		\multicolumn{1}{c}{\hyperlink{GD5}{GD5}}&0.46 M&0.49 M&0.49 M\\				
		\hline
		\multicolumn{1}{c}{\hyperlink{GD6}{GD6}}&0.61 H&0.57 M&0.57 M\\				
		\hline
		\multicolumn{1}{c}{\hyperlink{GD7}{GD7}}&0.40 L&0.43 M&0.60 H\\				
		\hline
		\multicolumn{1}{c}{\hyperlink{GD8}{GD8}}&0.31 L&0.36 L&0.51 M\\				
		\hline
		\multicolumn{1}{c}{\hyperlink{GD9}{GD9}}&0.53 M&0.51 M&0.71 H\\				
		\hline
		\multicolumn{1}{c}{\hyperlink{GD10}{GD10}}&0.24 L&0.31 L&0.51 M\\				
		\hline
		\multicolumn{1}{c}{\hyperlink{GD11}{GD11}}&0.50 M&0.54 M&0.70 H\\				
		\hline
		\multicolumn{1}{c}{\hyperlink{GD12}{GD12}}&0.55 M&0.46 M&0.60 H\\				
		\hline

	\end{tabular}
	\vspace{-0.4cm}

\end{table}
\subsubsection{Similarity Analysis}
As introduced previously, we analyzed the similarity between the focus areas of the FBSers and UBSers using the three indicators.

\par{The OS results indicated that \textbf{the background stories had a medium level of effect on the selection sequences of the participants}. The main reason was that the stories implied the existence of graph structures, specifically for high degree, bridge, and community structures. The FBSers may have formed preconceived notions when learning such existence information, and thus preferentially sought implied structures during graph explorations. We took \hyperlink{GD10}{GD10} (\autoref{fig:figure5}(o$  -  $p)) and \hyperlink{GD7}{GD7} (\autoref{fig:figure5}(i$  -  $j)) as examples. The story of \hyperlink{GD10}{GD10} implied the existence of high degree nodes (i.e., popular classmates) in a school class friendship network. We observed that the FBSers mainly sought high degree nodes in the central region of the graph, whereas the UBSers were largely attracted by peripheral nodes. The results of ${\textbf{\emph{OM}}_{FBSers,10}}$ and ${\textbf{\emph{OM}}_{UBSers,10}}$ showed that high degree structures received 25 and 7 entries of 1st from the FBSers and UBSers, respectively, whereas other structures received 5 and 17 entries of 1st from the FBSers and UBSers, respectively. Many of the UBSers commented, “\emph{No particular structures seem to be in the dense central area of the graph, but those nodes at the margin areas are interesting.}” The story of \hyperlink{GD7}{GD7} emphasized multilingual employees responsible for communication between groups that were defined based on language in a wood-processing facility, which implied the existence of bridge structures. We observed that most of the FBSers selected bridges first, whereas only nearly half of the UBSers initially selected bridges. The results of ${\textbf{\emph{OM}}_{FBSers,7}}$ and ${\textbf{\emph{OM}}_{UBSers,7}}$ showed that bridge structures received 26 and 14 entries of 1st from the FBSers and UBSers, respectively, whereas community structures received 8 and 20 entries of 1st from the FBSers and UBSers, respectively.}

\par{The SS results indicated that \textbf{the background stories moderately influenced the number of selecting specific graph structures.} The main reason was that some of the background stories directly or indirectly provided the number of graph structures that existed in graphs, which could have produced an anchoring effect \cite{A49} that drove the FBSers to subconsciously present a minimum number of specific graph structures to be selected in a graph. For example, the story of \hyperlink{GD3}{GD3} had two protagonists: the president and leading instructor of the Zachary’s karate club, which indicated that at least two high degree nodes existed in the graph. We noticed that many of the UBSers only selected one high degree node. The results of ${\textbf{\emph{SV}}_{FBSers,3}}$ and ${\textbf{\emph{SV}}_{UBSers,3}}$ showed that 2.29 and 1.00 high degree nodes were selected on average by each FBSer and UBSer, respectively. Such an anchoring effect also appeared when selecting bridges in \hyperlink{GD7}{GD7} and high degree nodes in \hyperlink{GD10}{GD10} and \hyperlink{GD12}{GD12}.}

\par{The PS results indicated that \textbf{the background stories had a medium level of effect on the focus areas of the participants regardless of specific graph structures}. We took \hyperlink{GD4}{GD4} and \hyperlink{GD5}{GD5}, which had relatively low PS scores, as examples to explain two findings in the PS results. The story of \hyperlink{GD4}{GD4} mentioned the slighted Baratheons and the exiled Daenerys in the \emph{Game of Thrones} novel. We found that the two small clusters far from the dense central area of the graph were concerned by the FBSers more than the UBSers, as shown in \autoref{fig:figure5}(c$  -  $d). This example reflected that the background stories could suggest the positions of important structures in the layout. The story of \hyperlink{GD5}{GD5} mainly described popular members in the TI baseball team. We found that many of the FBSers only selected high degree nodes, but the UBSers selected almost all nodes in the graph, as shown in \autoref{fig:figure5}(e$  -  $f). Some of the UBSers commented, “\emph{Each node is worth exploring in such a small-sized graph}.” This example reflected that the background stories may have bounded the participants’ focus areas.}

\subsection{Results of Experiment 2}
We assumed that background stories can affect the performance of identifying high degree, bridge, and community structures (H2). This hypothesis was partially confirmed. \textbf{The background stories were found to significantly affect community structure identifications ($\rho$ = 0.000  $<$ 0.05, Cohen's d = 0.521), but not high degree structure ($\rho$ = 0.644, Cohen's d = 0.039) and bridge structure ($\rho$ = 0.859, Cohen's d = 0.062) identifications in terms of accuracy.} We analyzed the quantitative and qualitative results of Ex2 by structure type (\autoref{tab:table3}) as follows. Detailed quantitative and qualitative results by objective question are provided in the supplementary materials.

\begin{table}[!ht]
	\scriptsize
	\vspace{0.15cm}
	\renewcommand\tabcolsep{5.3pt}  
	\renewcommand\arraystretch{1.5} 
	\centering
	\caption{ Quantitative and qualitative results of Ex2 by structure type in terms of mean accuracy, difficulty, and confidence level. The blue colors indicate significant differences between the FBSers and UBSers in terms of accuracy.}
	\label{tab:table3}

	\begin{tabular}{ccccccc}
		\hline  
		
		\multirow{2}{*}{\textbf{Structure Type}} &\multicolumn{2}{c}{\textbf{Accuracy}}&\multicolumn{2}{c}{\textbf{Difficulty}}&\multicolumn{2}{c}{\textbf{Confidence}}\\
		\cline{2-7}

		&\textbf{FBSers}& \textbf{UBSers}&\textbf{FBSers}& \textbf{UBSers}&\textbf{FBSers}& \textbf{UBSers}\\ 		
		\hline
		\multicolumn{1}{c}{High degree}&0.80&0.79&2.48&2.33&3.80&3.72\\	
		\hline
		\multicolumn{1}{c}{Bridge}&0.61&0.64&2.29&2.20&3.96&3.88\\				
		\hline
		\multicolumn{1}{c}{Community}&\cellcolor[rgb]{.33,.55,.84}\color{white}0.58&\cellcolor[rgb]{.55,.7,.88}\color{white}0.33&2.69&2.96&3.61&3.33\\				
		\hline
	\end{tabular}
	\vspace{-0.4cm}	

\end{table}

\subsubsection{Structure Identification Analysis}
We report the detailed analysis of comparing the participants’ performance on the identifications of three types of graph structures.

\par{\textbf{No significant difference was found in high degree structure identifications in terms of accuracy}. We had two observations in Ex2. The first exhibit showed that the participants estimated the degree of a node by assessing the number of edges centered on it, which could be considered as a numerical ground truth \cite{A22,A77}. The second exhibit showed that the 10 experimental graphs were small-sized and visualized without substantial visual clutter, which ensured that the numerical ground truth could be easily perceived. As a result, the information related to high degree structures in the background stories had minimal effects on accuracy. Moreover, we found that some of the background stories implied approximate locations of high degree structures in the node-link diagrams. These locations were perceptual “anchors” that facilitated structure identifications. For example, Q4 (\hyperlink{GD3}{GD3}) inquired two important high degree nodes in the member friendship network of the Zachary’s karate club. Some of the FBSers stated that “\emph{the two-part fission event of the club reminds me to seek high degree nodes on both sides of the graph}.”}
\par{\textbf{No significant difference was found in bridge identifications in terms of accuracy}. The mean accuracy of the FBSers was close to that of the UBSers. The reason was that bridges generally appeared in sparse boundary areas among dense communities. The participants can clearly observe bridges and visual community context for accurate identifications.}
\par{\textbf{The FBSers performed significantly better than the UBSers in identifying communities in terms of accuracy} ($\rho$ = 0.000, Cohen's d = 0.521), specifically for Q5, Q8, and Q12. The reasons were twofold. First, communities may overlap, which could have introduced visual ambiguities in community distinguishing \cite{A50}. Second, the community-related objective questions mainly inquired about the exact or approximate number of communities in a graph, and the corresponding background stories implied the answers. For Q5 (\hyperlink{GD3}{GD3}, $\rho$ = 0.018, Cohen's d = 0.588), as shown in \autoref{fig:figure5}(a$  -  $b), some of the FBSers stated, “\emph{The story describes the fission event of the club. Thus, I think there are two factions in the graph}.” For Q8 (\hyperlink{GD6}{GD6}, $\rho$ = 0.000, Cohen's d = 1.115), as shown in \autoref{fig:figure5}(g$  -  $h), some of the FBSers stated, “\emph{I have learned from the background information that each conference involves approximately 8 to 12 teams in American college football games. Thus, I can estimate the number of conferences by viewing the graph}.” For Q12 (\hyperlink{GD9}{GD9}, $\rho$ = 0.000, Cohen's d = 1.014), as shown in \autoref{fig:figure5}(m$  -  $n), many of the FBSers commented, “\emph{The story mentions that political ideologies mainly include ‘liberal’, ‘neutral’, and ‘conservative’. Thus, the co-purchasing network of political books should have three groups}.” Moreover, we noticed that the stories of \hyperlink{GD4}{GD4} and \hyperlink{GD7}{GD7} exerted a relatively low influence on community identifications because the community information in these stories was obscure or the communities were clearly presented without visual ambiguities. Some of the FBSers suggested that the community structure of the \emph{Game of Thrones} character relationship network (\hyperlink{GD4}{GD4}) was complex, as shown in \autoref{fig:figure5}(c$  -  $d), and the names of a few families provided in the story had limited usefulness in judging the number of families. Some of the UBSers commented that Q9 was the easiest among all community-related questions because the three communities in the wood-processing facility (\hyperlink{GD7}{GD7}) were well-separated in the node-link diagram, as shown in \autoref{fig:figure5}(i$  -  $j).}
\subsubsection{Participants’ Ratings}
\par{The qualitative results of Ex2 were consistent with the quantitative results. The mean ratings of the FBSers and UBSers were close in high degree and bridge identifications in terms of difficulty and confidence levels ranging from 1 (lowest level) to 5 (highest level). The mean difficulty rating of the FBSers was lower than that of the UBSers and the mean confidence rating of the FBSers was higher than that of UBSers in community identifications. Moreover, communities obtained the highest mean difficulty rating and the lowest mean confidence rating among the three structure types, which indicated that the background stories helped complete relatively difficult tasks. Specific to individual objective questions, Q6 (\hyperlink{GD4}{GD4}) and Q8 (\hyperlink{GD6}{GD6}) received many high difficulty ratings $\left ( \mu > 3 \right )$ because they were community-related questions. Q13 (\hyperlink{GD10}{GD10}), which inquired about high degree structures, obtained a mean rating ($\mu$ = 3.37) of high difficulty because more than 10 high degree nodes were concentrated in the central dense area of the graph, as shown in \autoref{fig:figure5}(o$  -  $p).}

\subsection{Results of Experiment 3}
We assumed that background stories can help viewers construct stable mental models for graph recognition under difficult visual conditions (H3). This hypothesis was fully confirmed. \textbf{The FBSers performed significantly better than the UBSers in graph recognition in terms of accuracy ($\rho$ = 0.000 $<$ 0.05, Cohen's d = 0.139).} The mean accuracy of 40 trials of the FBSers ($\mu$ = 0.76) was slightly higher than that of the UBSers ($\mu$ = 0.70). Specific to individual graphs, significant differences were found in recognizing \hyperlink{GD4}{GD4} ($\rho$ = 0.001, Cohen's d = 0.387), \hyperlink{GD5}{GD5} ($\rho$ = 0.013, Cohen's d = 0.301), and \hyperlink{GD12}{GD12} ($\rho$ = 0.008, Cohen's d = 0.322), as shown in \autoref{tab:table4}.

\begin{table}[!t]
	\scriptsize
	\renewcommand\tabcolsep{6.35pt}  
	\renewcommand\arraystretch{1.6} 
	\centering
	\vspace{0.15cm}  
	
	\caption{ Quantitative and qualitative results of Ex3 by graph (4 trials per graph) in terms of mean accuracy, difficulty, and confidence level. The blue colors indicate significant differences between the FBSers and UBSers in terms of accuracy.}
	\label{tab:table4}

	\begin{tabular}{ccccccc}
		\hline  
		
		\multirow{2}{*}{\textbf{Graph ID}} &\multicolumn{2}{c}{\textbf{Accuracy}}&\multicolumn{2}{c}{\textbf{Difficulty}}&\multicolumn{2}{c}{\textbf{Confidence}}\\
		\cline{2-7}

		&\textbf{FBSers}& \textbf{UBSers}&\textbf{FBSers}& \textbf{UBSers}&\textbf{FBSers}& \textbf{UBSers}\\ 		
		\hline
		\multicolumn{1}{c}{\hyperlink{GD3}{GD3}}&0.68&0.63&2.58&2.72&3.51&3.28\\	
		\hline
		\multicolumn{1}{c}{\hyperlink{GD4}{GD4}}&\cellcolor[rgb]{.33,.55,.84}\color{white}0.84&\cellcolor[rgb]{.55,.7,.88}\color{white}0.67&2.34&2.55&3.76&3.54\\				
		\hline
		\multicolumn{1}{c}{\hyperlink{GD5}{GD5}}&\cellcolor[rgb]{.33,.55,.84}\color{white}0.88&\cellcolor[rgb]{.55,.7,.88}\color{white}0.76&2.28&2.41&3.96&3.60\\		
		\hline
		\multicolumn{1}{c}{\hyperlink{GD6}{GD6}}&0.72&0.72&2.41&2.55&3.72&3.54\\
		\hline
		\multicolumn{1}{c}{\hyperlink{GD7}{GD7}}&0.86&0.80&2.07&2.33&4.07&3.78\\\hline
		\multicolumn{1}{c}{\hyperlink{GD8}{GD8}}&0.78&0.69&2.24&2.44&3.82&3.58\\\hline
		\multicolumn{1}{c}{\hyperlink{GD9}{GD9}}&0.67&0.69&2.68&2.70&3.58&3.33\\\hline
		\multicolumn{1}{c}{\hyperlink{GD10}{GD10}}&0.64&0.70&2.51&2.52&3.57&3.44\\\hline
		\multicolumn{1}{c}{\hyperlink{GD11}{GD11}}&0.76&0.69&2.24&2.21&3.89&3.77\\\hline
		\multicolumn{1}{c}{\hyperlink{GD12}{GD12}}&\cellcolor[rgb]{.33,.55,.84}\color{white}0.79&\cellcolor[rgb]{.55,.7,.88}\color{white}0.65&2.19&2.29&3.85&3.68\\		
		\hline
		
	\end{tabular}
	\vspace{-0.4cm}	

\end{table}

\subsubsection{Quantitative Analysis}
\par{We obtained two important findings from the quantitative results of Ex3. The first finding was that \textbf{the background stories helped the FBSers learn immutable visual features}, including symmetry, collinearity, and orthogonality \cite{A7}, from the graph visualizations. There were three facets to these visual features. First, they were susceptible to interference from blurring and flipping. Second, perfectly preserving them in generated interference graphs was difficult. Third, some of the background stories implied their existence. Therefore, the FBSers were able to easily notice them and consciously or subconsciously internalized them as graph representations to construct stable mental models for graph perception. For example, many of the FBSers confirmed that the slighted Baratheons and the exiled Daenerys in the \emph{Game of Thrones} novel (\hyperlink{GD4}{GD4}) were two small clusters far from the dense central area of the graph, and they were symmetrically distributed. Some of the FBSers commented, “\emph{The background information of the SA’s baseball team describes three key players, and I find the three players are almost collinear in the central area of the graph. This feature is very helpful in graph recognition} (\hyperlink{GD12}{GD12} shown in \autoref{fig:figure5}(s$  -  $t)).” A few of the FBSers said, “\emph{I notice several orthogonal edges converging on an important person mentioned in the background story. I use this pattern to distinguish this graph from others} (\hyperlink{GD11}{GD11} shown in \autoref{fig:figure5}(q$  -  $r)).”}
\begin{figure*}[!ht]
	\centering
	\vspace{-0.03cm}  
	\setlength{\abovecaptionskip}{0.1cm}   
	\setlength{\belowcaptionskip}{-0.6cm}
	\includegraphics[width=18cm]{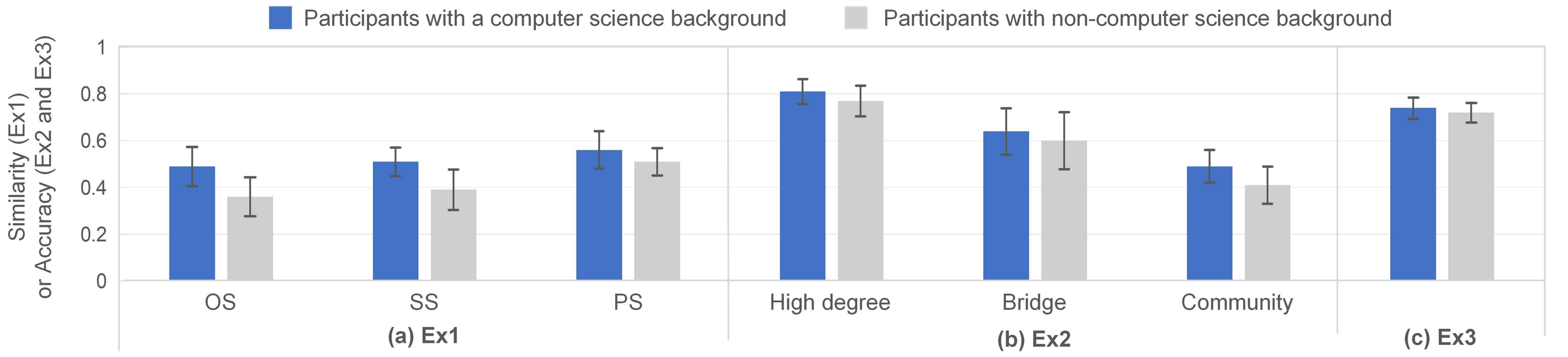}
	\caption{Results on the influence of participants’ professional backgrounds: (a) mean similarity values of the three similarity indicators in Ex1; (b) mean accuracy values when identifying high degree nodes, bridges, and communities in Ex2 and (c) mean accuracy values in Ex3. Error bars show 95\% confidence intervals.}
	\label{fig:figure6}
\end{figure*}
\par{The second finding was that \textbf{the background stories simulated outline associations of the FBSers}. In outline associations, viewers build associations between the outlines of perceived visual stimuli and those of familiar objects \cite{A7}. For example, some of the FBSers suggested that the outlines of \hyperlink{GD4}{GD4}, \hyperlink{GD5}{GD5}, and \hyperlink{GD6}{GD6} resembled a person raising both hands, a whale, and the head of a pig, respectively. The graph mental models that were formatted based on outline associations were not easy to remove or to be affected by blurring, and interference graphs rarely presented consistent outlines with reference raw graphs. Interestingly, such outline associations were mainly reported by the FBSers, and no aforementioned associated objects were described in the background stories. This situation indicated that the background stories played the role of stimulating associations rather than straightforwardly telling the FBSers what outlines looked like.}

\par{Moreover, the mean accuracies of the FBSers were lower than those of the UBSers when recognizing \hyperlink{GD9}{GD9} and \hyperlink{GD10}{GD10}. For \hyperlink{GD9}{GD9} (\autoref{fig:figure5}(m$  -  $n)), the graph presented an overall shape with two visual communities, which was not consistent with the ground truth of three communities mentioned in the background story. For \hyperlink{GD10}{GD10} (\autoref{fig:figure5}(o$  -  $p)), most of the participants reported that no prominent visual features existed in the graph except for a ball-like cluster in the central area. Thus, the circumstances of recognitive ineffectiveness of the background stories were twofold. The information in a background story had poor correspondence with graph structures or visual features. A graph had no noticeable visual features.}

\subsubsection{Participants’ Ratings}
The qualitative results of Ex3 were consistent with the quantitative results. The mean difficulty rating of the FBSers ($\mu$ = 2.35) was slightly lower than that of the UBSers ($\mu$ = 2.47), and the mean confidence rating of the FBSers ($\mu$ = 3.77) was higher than that of the UBSers ($\mu$ = 3.55). For individual graphs, \hyperlink{GD7}{GD7} (\autoref{fig:figure5}(i$  -  $j)) obtained a relatively low mean difficulty rating because of small size and apparent graph structures. \hyperlink{GD9}{GD9} (\autoref{fig:figure5}(m$  -  $n)) obtained a relatively high mean difficulty rating due to the absence of noticeable visual features. Moreover, when asked about the three difficult conditions in the interviews, many of the participants suggested that the blurred interference graphs affected graph recognition more than the blurred and flipped raw graphs, which affected graph recognition more than the blurred raw graphs.


\subsection{Influence of Participants' Backgrounds}
We were interested in whether the participants' professional backgrounds had an influence on graph perception. We divided the participants into two groups named computer-science group (40 participants) and non-computer-science group (30 participants) based on their professional backgrounds. We re-analyzed the raw results of the three experiments according to the two groups. Main results are shown in \autoref{fig:figure6} and detailed results are provided in the supplementary materials.
\par{In Ex1, \textbf{the mean values of the three similarity indicators of the computer-science group (OS: $\mu$ = 0.49; SS: $\mu$ = 0.51; PS: $\mu$ = 0.56) were larger than those of the non-computer-science group (OS: $\mu$ = 0.36; SS: $\mu$ = 0.39; PS: $\mu$ = 0.51)} and the mean OS and SS values of the non-computer-science group fell into a low similarity level [0.0, 0.4]. The reasons were twofold. First, most of the computer science participants had learned relevant subjects, such as data structures, graph theory, or complex networks, and grasped the elementary knowledge of typical graph structures (e.g., high degree nodes, bridges, and communities), which led them to give priority to these structures. Taking \hyperlink{GD10}{GD10} (\autoref{fig:figure5}(o$  -  $p)) as an example, 89 and 32 high degree nodes were selected in total by the computer-science and non-computer-science groups, respectively. Many participants in the computer-science group focused on high degree nodes at the very beginning due to having knowledge that high degree nodes play an important role in a graph. Second, the participants in the non-computer-science group had diverse professional backgrounds, which may have lead to their differentiated preferences in selecting AOIs. For example, the story of \hyperlink{GD8}{GD8} described the relationship between different states and legal bases for divorce. Some participants in the non-computer-science group knew much about the United States from state to state while others were interested in divorce laws, leading to differences in their AOI selections, as shown in \autoref{tab:table2} and (\autoref{fig:figure5}(k$  -  $l))).
}
\par{In Ex2, \textbf{the mean accuracy values of the computer-science group when identifying the three types of structures (high degree, $\mu$ = 0.81; bridge, $\mu$ = 0.64; community, $\mu$ = 0.49) were higher than those of the non-computer-science group (high degree, $\mu$ = 0.77; bridge, $\mu$ = 0.60; community, $\mu$ = 0.41).} No significant difference was found between the two groups in terms of accuracy (high degree: $\rho$= 0.345 $>$ 0.05, Cohen’s d = 0.102; bridge: $\rho$ = 0.157 $>$ 0.05, Cohen’s d = 0.071; community: $\rho$ = 0.146 $>$ 0.05, Cohen’s d = 0.157), but a relatively large difference was found when identifying communities (computer-science group: $\mu$ = 0.49; non-computer-science group: $\mu$ = 0.41).
	
}
\par{In Ex3, {\textbf{the computer-science ($\mu$ = 0.74) and non-computer-science ($\mu$ = 0.72) groups obtained approximate mean accuracy values in graph recognition under difficult visual conditions.}} No significant difference was found between the two groups in terms of accuracy ($\rho$ = 0.499 $>$ 0.05, Cohen's d = 0.026). Feedback from the participants showed that they mainly adopted the strategy of outline associations for graph memorization and recognition. For example, some of the participants imagined \hyperlink{GD3}{GD3} as a sailing boat and \hyperlink{GD4}{GD4} as a person raising both hands. Such outline associations were slightly related to the specialized knowledge of typical graph structures but largely based on perceptions of common objects in daily life.
}

\subsection{Summary of Results}
The results of Ex1 indicated that the background stories affected the sequences, numbers, and positions of AOI selections made by the FBSers. The reason was twofold. The background information about the existence and positions of graph structures formed preconceived notions. Thus, the FBSers preferentially selected preconceived structures. The number of graph structures implied in the background stories produced an anchoring effect that drove the FBSers to subconsciously present a minimum number of selecting specific structures.
\par{The results of Ex2 reflected that the background stories significantly affected community identifications in terms of accuracy because many communities in the experimental graphs overlapped, and the UBSers had difficulty distinguishing them without knowing the relevant information in the background stories. However, the stories had minimal effects when identifying high degree and bridge structures because the small-sized experimental graphs caused little visual clutter. Thus, the participants can perceive them by using the numerical ground truth and displayed community context.}
\par{The results of Ex3 confirmed that the background stories helped the FBSers construct stable mental models to perform accurate graph recognition under difficult visual conditions. The background stories facilitated the FBSers to learn immutable visual features (i.e., symmetry, collinearity, and orthogonality) and simulated their outline associations with familiar objects. Immutable visual features and outline associations created stable graph representations that helped with mental model formation.}
\par{The results of influence analysis of professional backgrounds indicated that the participants' backgrounds lead to different AOI selections during open-ended graph explorations (Ex1). There was no significant difference in the performance of identifying the three types of structures (Ex2) and graph recognition under difficult visual conditions (Ex3).
}

\section{Discussion}	
In this section, we discuss the limitations of this study, provide implications for visualization design, and suggest extensions for future work.
\subsection{Limitations}
We discuss the limitations of this study from the aspects of data, experimental design, and results.
\par{The graph data sets used in the experiments were not extensive. We initially collected 38 candidates with diverse types and sizes, but only one-third of them met all the four selection criteria. Specialized graphs, such as biological networks \cite{A51} and autonomous system networks \cite{A52}, were difficult for our participants to understand. Large-scale graphs caused visual clutter to hinder Ex1 and Ex2. Some candidates presented similar overall shapes that were inadaptable for Ex3. Three possible directions can support the expansion of experimental graph data sets: (1) conducting the three experiments separately so that mid-sized graphs can be used for Ex1 and Ex3, (2) providing zooming or fisheye interactions for mid-sized graphs in Ex2, and (3) inviting domain experts to analyze specialized graphs.}
\par{The scope of bridge structures in this study was not as strict as that in graph theory. In graph theory, a bridge is an edge whose removal increases the number of connected components of a graph \cite{A53}. In this study, we considered nodes and edges that connect any two groups/communities as bridges. This was done for two reasons. First, our scope was close to the descriptions in the background stories. Second, theoretical bridges were not prevalent in the experimental graphs. The latter reason also resulted in a small number of bridge-related objective questions in Ex2. The number of participants in this study was limited. Although we controlled the number of graphs and questions, the entire set of experiments still lasted at least three hours. We did not use crowdsourcing techniques or platforms, such as Amazon Mechanical Turk \cite{A54}, to increase the size of our participant pool because a previous study found that crowdworkers may not be particularly interested in carefully reading the provided narratives and may prefer to quickly complete their tasks instead \cite{A87}. This behavior would not have been suitable for our experimental design because our design required close supervisions due to the similarities between experiments and relatively complicated tasks. Therefore, it was not clear what the results would be if a large number of volunteers participate in the experiments.}
\par{This study investigated the short-term memory but not the long-term memory of graphs \cite{A7} because the participants started the testing session immediately after the learning session in Ex3. This study adopted two types of difficult visual conditions in Ex3, namely, mirroring and blurring, but did not investigate their mechanisms to influence graph memorization or recall. This study did not evaluate the effect of background stories on increasing the engagement of the participants during interactive graph explorations. A previous study \cite{A88} suggested that the engagement was an interesting topic in storytelling visualizations and can be defined from either a behavioral or an emotional perspective. This study used the same graph layout, but a previous study showed that perceived structures were different as the spatial arrangement of a graph changes \cite{A55}. Thus, we were unsure what the experimental results would be with other layouts. This study recorded the duration results but did not provide a detailed result analysis of duration. Pilot studies showed that some participants were slightly hesitant sometimes to select specific nodes/structures in some graphs (Ex1 and Ex2) or determine answers under difficult visual conditions (Ex3). We decided to allow answer modifications before proceeding to the next graph (Ex1), objective question (Ex2), or trial (Ex3) without time limits in the formal experiment. Thus, large differences in time consumption were observed among the participants in the three experiments.}
\subsection{Implications}
This research provides design implications for graph storytelling and interactive graph explorations.
\par{Learning from the results of Ex1, using background stories can be to guide the attention of viewers and stimulate their interest while exploring target graph structures, such as high degree, bridge, or community structures, because not all structures in a graph are of equal importance in a given analysis scenario and the focus areas of viewers can be distracted by trivial structures. For example, if we expect viewers to pay attention to high degree nodes in a graph, we should explicitly describe the existence of several central characters in the background story of the graph.}
\par{The results of Ex2 indicate that the background story of a graph as a visual annotation is crucial during interactive graph explorations, especially for community structure analysis. This is because visual ambiguities of community structures are generally larger than those of high degree and bridge structures in node-link diagrams. For example, to help viewers accurately identify communities in the visualization of a graph, we can provide information about the number or size of a community structure in the background story.
}
\par{The results of Ex3 suggest that providing background stories can facilitate dynamic graph analysis \cite{A60,A61} and graph comparison analysis \cite{A62,A63}. This is because stable graph mental models can help viewers perceive the changes of overall shapes or important structures in graph visualizations. For example, if the background story of a dynamic graph mentions that the characters connecting two communities disappear after an important event, the change of bridging structures in graph visualizations will be easily noticed by viewers.}
\par{Taking all the results together, the design of graph storytelling should consider the concurrent display and collaborative design of narrative texts and graph visualizations to create a synergy effect \cite{A56,A78}. The concurrent display refers to the presentation of narrative texts together with graph visualizations on the screen. Texts convey the real-world meanings of entities, relations, and structures in graphs. Visualizations, on the other hand, provide supporting evidence and relevant details \cite{A57}, which can improve viewers' comprehension and ability to recognize informaiton \cite{A58,A59}. The collaborative design refers to the iterative refinement of displayed texts and visual encodings to ensure the dual-way confirmation of textual and visual information. A good narrative design is as important as a good visualization design \cite{A87,A91}. That is, the design of visual encodings should consider that narrative texts can affect the observing orders and focus areas of viewers. Narrative texts should be refined according to observable visual patterns in node-link diagrams. Moreover, the design of graph storytelling should consider the viewers' professional backgrounds, which can moderately influence their visual perceptions of graph visualizations.}

\section{Conclusion and Future Work}
This research evaluated the effects of background stories on graph perception. The experimental findings demonstrated that background stories can affect the focus areas of viewers. It also found that background stories can significantly affect community identification and graph recognition. This work is the first attempt that evaluates the effects of background stories on graph perception. The findings may bring new insights into the inherent laws of graph perception and new considerations about the design of storytelling visualizations and interactive graph explorations.
\par{Many future extensions can be conducted. First, a wide range of graphs and participants should be included to understand the generalizability of the findings. Second, eye-tracking techniques \cite{A65,A72} could be used in Ex1 to record more specific metrics (e.g., fixation duration \cite{A66}) that allude to areas of interest. Third, additional experimental conditions, such as including other graph structures (i.e., margin structures) in Ex2 and more difficult visual conditions (i.e., rotation \cite{A67} and size reduction \cite{A68}) in Ex3, should be considered. Finally, while the effects of background stories may improve the long-term memory of graphs and the engagement of users, this needs to be explicitly confirmed by future studies. Moreover, we hope that this study will inspire researchers to further investigate the effects of other personal knowledge and experience on graph perception. We also expect that the effects of background stories on visual perceptions of abstract data models, such as trees, tabulations, and trajectories, will be evaluated in the future.
}

\ifCLASSOPTIONcaptionsoff
  \newpage
\fi

\bibliographystyle{abbrv-doi}
\bibliography{myreference-doi}

\end{document}